\documentclass[preprint,showpacs,preprintnumbers,amsmath,amssymb]{revtex4}
\usepackage{amsmath}
\usepackage{amssymb}
\usepackage{graphicx}
\begin{document}
\title{Mapping from current densities to vector potentials in time-dependent current-density functional theory}
\author{Giovanni Vignale}
\affiliation{Department of Physics and Astronomy, University of
Missouri - Columbia, 65211, Columbia, Missouri, USA}
\begin{abstract}
We show  that the time-dependent particle density $n(\vec r,t)$ and the current density ${\vec j}(\vec r,t)$ of a many-particle system that evolves under the action of external scalar and vector potentials $V(\vec r,t)$ and $\vec A(\vec r,t)$ and is initially in the quantum state $|\psi (0)\rangle$, can always be reproduced (under mild assumptions) in another many-particle system, with different two-particle interaction, subjected to external potentials  $V'(\vec r,t)$ and $\vec A'(\vec r,t)$, starting from an initial state $|\psi' (0)\rangle$, which yields the same density and current as $|\psi (0)\rangle$.   Given the initial state of this other many-particle system, the potentials $V'(\vec r,t)$ and $\vec A'(\vec r,t)$ are uniquely determined up to gauge transformations that do not alter the initial state.   As a special case, we obtain a new and simpler proof of the Runge-Gross theorem for time-dependent current density functional theory.   This theorem provides a formal basis for the application of time-dependent current density functional theory to transport problems.
\end{abstract}
\date{\today}
%\showpacs{}
\maketitle
\section{Introduction}
Time-dependent density functional theory (TDDFT) attempts to describe the influence of many-body interactions on the the time evolution of a quantum many-particle system in terms of an effective local potential that depends on a single collective variable, the particle density $n(\vec r,t)$.  The possibility of  such a description is largely based on a fundamental theorem,  proved by Runge and Gross~\cite{Runge84} in 1984, which guarantees the invertibility of the mapping from time-dependent potentials $V(\vec r,t)$ to time-dependent densities, for time-evolutions that start from a common initial state  $|\psi(0)\rangle$.  More precisely, this theorem asserts that two different potentials $V(\vec r,t)$ and $V'(\vec r,t)$ which are both analytic functions of $t$ in a neighborhood of the initial time $t=0$, and which are {\it not} trivially related by a gauge transformation (i.e., by the addition of a merely time-dependent constant to the scalar potential), cannot give rise to the same density, starting from the same initial state:  therefore $n(\vec r,t)$ determines $V(\vec r,t)$ uniquely up to a gauge transformation~\cite{footnote}.

The Runge-Gross theorem has been considerably strengthened a few years ago by the proof of another theorem, which we refer to as the {\it van Leeuwen's theorem}, which states  that the time-dependent particle density  of a many-particle system that evolves under the action of an external  potential  $V(\vec r,t)$ and is initially in the state $|\psi (0)\rangle$, can always be reproduced (under mild assumptions) in another many-particle system, with different two-particle interaction, subjected to an external potential  $V'(\vec r,t)$,  starting from an initial state $|\psi' (0)\rangle$ which yields the same density and divergence of the current density as $|\psi (0)\rangle$.   Given the initial state of this other many-particle system, the potential $V'(\vec r,t)$ is uniquely defined up to a gauge transformation.      The content of the van Leeuwen theorem is particularly interesting in two special cases:   (i) if the second many-particle system is non-interacting, then the theorem guarantees the possibility of reproducing the time evolution of the density of an interacting many-particle system in a non-interacting many-particle system,  (ii) if the second many-particle system has the same two-particle interaction as the first, then the theorem becomes equivalent to the Runge Gross theorem, which is thereby proved in a more direct manner.

Although the TDDFT has been remarkably successful over the 20 years of its history,  there are several situations in which it seems useful to reformulate the theory in terms of the particle current density $\vec j(\vec r,t)$, leading to what is known as time-dependent current density functional theory (TDCDFT)~\cite{VK97}.   It has been pointed out that the time-dependent exchange-correlation potential, when regarded as a functional of the density,  does not admit a gradient expansion~\cite{VK97} (this is the so-called ``ultra non-locality problem"), whereas the exchange-correlation vector potential  of TDCDFT does. 
For this reason there has been great interest in applying TDCDFT to situations in which the standard TDDFT is known to have problems, such as optical spectra of solids~\cite{deBoeij02,Maitra03} and  dielectric properties of polymer chains~\cite{vanFaassen}. The reformulation of TDDFT in terms of the current density is particularly attractive  to those who wish to apply DFT methods to molecular transport problems~\cite{DiVentra02}, for  the TDCDFT gives direct access to the electrical current density.   Finally, it should be noted that this theory allows one to dispose of boundary conditions at infinity, such as the ones mentioned in footnote~\cite{footnote}. 

Efforts to provide a rigorous basis for TDCDFT date back to the pioneering work of Ghosh and Dhara~\cite{Ghosh88}, where the invertibility (up to a gauge  transformation) of the map from scalar and vector potentials $V(\vec r,t),\vec A(\vec r,t)$ to particle density and current density $n(\vec r,t),\vec j(\vec r,t)$  was first proved under hypotheses similar to those of the original Runge-Gross theorem.  The Ghosh-Dhara proof  of this theorem is considerably more complex than the  original proof of the Runge-Gross theorem and, furthermore,  does not address the issue of representability of the interacting current density evolution in a non-interacting system, which is central to the foundation of the Kohn-Sham equation. In this paper we overcome both limitations by proving the analogue of van Leeuwen's theorem in TDCDFT.  The precise statement of the theorem is as follows:

{\bf Theorem.}  Consider a many-particle system described by the time-dependent hamiltonian
\begin{equation}\label{H}
\hat H (t) =\sum_{i} \left[\frac{1}{2m}\left (\hat {\vec p}_i + \vec  A(\hat {\vec r}_i,t)\right)^2 +V(\hat {\vec r}_i,t) \right]+ \sum_{i<j} U(\hat {\vec r}_i,\hat {\vec r}_j)~,
\end{equation}
where $V(\vec r,t)$ and $\vec A(\vec r,t)$ are given external scalar and vector potentials, which are analytic functions of time in a neighborhood of $t=0$, and $U$ is a two-particle interaction.  Let  $n(\vec r,t)$ and $\vec j(\vec r,t)$  be the particle density and current density  that evolve under $\hat H$ from a given initial state $|\psi(0)\rangle$.   Then the same density and current density can be obtained from another many-particle system, with Hamiltonian
\begin{equation}\label{Hprime}
\hat H^\prime (t) =\sum_{i} \left[\frac{1}{2m}\left (\hat {\vec p}_i + \vec  A^\prime(\hat {\vec r}_i,t)\right)^2 +V^\prime(\hat {\vec r}_i,t) \right]+ \sum_{i<j} U^\prime(\hat {\vec r}_i,\hat {\vec r}_j)~,
\end{equation}
starting from an initial state $|\psi'(0)\rangle$ which yields the same density and current density as  $|\psi(0)\rangle$ at time $t=0$.  The potentials $V'(\vec r,t)$ and $\vec A'(\vec r,t)$ are uniquely determined  by $V(\vec r,t)$, $\vec A(\vec r,t)$, $|\psi(0)\rangle$, and $|\psi'(0)\rangle$, up to gauge transformations of the form
\begin{eqnarray}\label{gaugetransformation}
V'(\vec r,t) &\to& V'(\vec r,t)-\frac{\partial \Lambda (\vec r,t)}{\partial t}\nonumber\\
\vec A'(\vec r,t) &\to& \vec A'(r,t)+\vec \nabla \Lambda (\vec r,t)~,
\end{eqnarray}
where $\Lambda$ is an arbitrary regular function of $\vec r$ and $t$,   which satisfies the initial condition $\Lambda(\vec r,0)=0$.

The proof of this theorem is presented in the next section, and its physical implications are discussed in Section III.

\section{Proof}
As a first step toward the proof of our theorem let us observe that, given a set of potentials $V(\vec r,t),\vec A(\vec r,t)$, one can always make a gauge transformation of the form~(\ref{gaugetransformation}) that kills the scalar potential at all times.   To this end, one simply chooses $\Lambda(\vec r,t)$ to be the solution of the differential equation 
\begin{equation} \label{radiationgauge}
\frac{\partial \Lambda (\vec r,t)}{\partial t}= V(\vec r,t)
\end{equation}
with initial condition $\Lambda(\vec r,0)=0$.  
We will henceforth assume that such a transformation has been done in both the unprimed and primed systems so that the scalar potentials $V$ and $V'$ are zero at all times. 
 
The current density operator in the unprimed system is given by
\begin{equation}\label{currentoperator}
\hat {\vec j}(\vec r,t) = \frac{1}{2}\sum_i \left \{\hat {\vec v}_i (t), \delta (\vec r - \hat{\vec r}_i) \right\}~,
\end{equation}
where $\{\hat A,\hat B\} \equiv (\hat A \hat B+\hat B \hat A)$ denotes the {\it anticommutator} of two operators $\hat A$ and $\hat B$, and the velocity operator of the $i$-th particle is given by
\begin{equation}\label{velocityoperator}
\hat {\vec v}_i = \frac{1}{m}\left[\hat{\vec p}_i + \vec A(\hat{\vec r}_i,t)\right]
\end{equation} 	
where $\hat {\vec p}_i = - i  \hbar \frac{\partial}{\partial \hat{\vec r}_i}$ is the canonical momentum operator.  Notice that this depends explicitly on time via the vector potential.  A completely analogous expression can be written, of course, for the current density of the primed system, the only difference being the replacement of the velocity operator $\hat {\vec v}_i$ by $\hat {\vec v}_i^\prime$, which is given by Eq.~(\ref{velocityoperator}) with $\vec A$ replaced by $\vec A'$.  

Let us denote by $\vec j(\vec r,t)$ and $n(\vec r,t)$ the expectation values of the current density and density operators in the quantum state that evolves from the initial state $|\psi(0)\rangle$ under the hamiltonian $\hat H$.  It is not difficult to verify that $\vec j(\vec r,t)$ obeys the equation of motion 
\begin{eqnarray}\label{eom}
\frac{d\vec j(\vec r,t)}{dt} &=& \frac{\partial \vec j(\vec r,t)}{\partial t} +\frac{i}{\hbar} \langle [\hat H(t),\hat {\vec j}(\vec r,t)]\rangle \nonumber \\ &=& n(\vec r,t) \frac{\partial \vec A(\vec r,t)}{\partial t} -  \vec j(\vec r,t) \times \left [\vec \nabla \times \vec A(\vec r,t)\right] +  {\bf \nabla}\cdot \langle{\bf \hat \sigma}(\vec r,t)\rangle~,
\end{eqnarray}
where  $\frac{\partial \vec A(\vec r,t)}{\partial t}$ and  $\vec \nabla \times \vec A(\vec r,t)$ are, respectively, the electric and the magnetic field, ${\bf \hat \sigma}$ is a  stress tensor operator,  defined below, and $\langle ... \rangle$ denotes the average in the quantum state of the unprimed system  at time $t$.   The  stress tensor is explicitly defined as 
\begin{eqnarray}\label{stresstensor}
\hat \sigma_{\alpha \beta}(\vec r,t) = -\frac{1}{4}\sum_i \left\{\hat v_\alpha, \left\{\hat v_\beta, \delta(\vec r - \hat {\vec r}_i)\right\}\right\}-\sum_i\sum_{j\neq i} U(\hat{\vec r}_i, \hat{\vec r}_j) \delta(\vec r - \hat{\vec r}_i) \delta_{\alpha \beta}~,
\end{eqnarray}
and its ``divergence" is a vector with components $[{\bf \nabla}\cdot{\bf \hat \sigma}(\vec r,t)]_\alpha = \sum_\beta \frac{\partial \hat \sigma(\vec r,t)}{\partial r_\beta}$.

By hypothesis, the current density also obeys the equation 
\begin{eqnarray}\label{eom2}
\frac{\partial \vec j(\vec r,t)}{\partial t}~=~ n(\vec r,t) \frac{\partial \vec A'(\vec r,t)}{\partial t} -  \vec j(\vec r,t) \times \left [\vec \nabla \times \vec A'(\vec r,t)\right] +  {\bf \nabla} \cdot\langle {\bf \hat \sigma'}(\vec r,t)\rangle^\prime~,
\end{eqnarray}
where $\langle ... \rangle^\prime$ denotes the average in the quantum state of the primed system  at time $t$.  Notice that the stress tensor ${\bf \hat \sigma'}$ differs from  ${\bf \hat \sigma}$  in two ways:  first, because it contains the velocity operator $\hat{\vec v}^\prime$ instead of  $\hat{\vec v}$ and, second, because it contains  the two-particle interaction $U'$ instead of $U$.   Taking the difference of the two equations~(\ref{eom}) and~(\ref{eom2})  we get
\begin{eqnarray}\label{keyequation1}
n(\vec r,t) \frac{\partial  \Delta \vec A (\vec r,t)}{\partial t}&=& \vec j(\vec r,t) \times \left [\vec \nabla \times \Delta \vec A(\vec r,t)\right] - {\bf \nabla}\cdot \langle{\bf \hat \sigma'}(\vec r,t)\rangle^\prime + {\bf \nabla} \cdot \langle{\bf \hat \sigma}(\vec r,t)\rangle ~,
\end{eqnarray}
where $\Delta \vec A (\vec r,t) \equiv \vec A'(\vec r,t) - \vec A(\vec r,t)$.  
Since, by hypothesis, both $\vec A(\vec r,t)$ and $\vec A'(\vec r,t)$  are expandable in a Taylor series of time in a neighborhood of $t=0$,  it follows that their difference, $\Delta \vec A (\vec r,t)$ is Taylor-expandable too.  We can therefore write  
\begin{equation}
\Delta \vec A(\vec r,t) = \sum_{k=0}^{\infty}\Delta A_k(\vec r) t^k~,
\end{equation}
where $\Delta A_k(\vec r)\equiv\frac{1}{k!}\left.\frac{\partial^k  \Delta \vec A(\vec r,t)}{\partial t^k}\right\vert_{t=0}$.
Substituting this expansion into Eq.~(\ref{keyequation1}) and equating the $l$-th term  of the Taylor expansion on each side of it we easily arrive at
\begin{eqnarray}\label{keyequation2}
\sum_{k=0}^{l}n_{l-k}(\vec r) \left[\frac{\partial  \Delta \vec A (\vec r,t)}{\partial t}\right]_{k} &=&  \sum_{k=0}^{l}\left \{\vec j_{l-k}(\vec r)\times \left[\vec \nabla \times \Delta \vec A_k(\vec r)\right] \right\}\nonumber \\ &+& [{\bf \nabla}\cdot \langle{\bf \hat \sigma}(\vec r,t)\rangle^\prime]_l - [{\bf \nabla} \cdot \langle{\bf \hat \sigma'}(\vec r,t)\rangle]_l~,\nonumber\\
\end{eqnarray}
where $n_{k}(\vec r)$ and $\vec j_k(\vec r)$ denote the $k$-th coefficients in the Taylor expansions of $n(\vec r)$ and $\vec j(\vec r)$ about $t=0$,   and, more generally  $[f(t)]_l$ denotes the $l$-th coefficient in the expansion of a function $f(t)$ in powers of $t$ about $t=0$.  The fact that all the quantities appearing in the above equation admit such an expansion is a consequence of the analyticity of the vector potential and of the time-dependent Schr\"odinger equation
\begin{equation}\label{SE}
i\hbar \frac{\partial |\psi(t)\rangle}{\partial t}=\hat H(t) |\psi(t)\rangle~.
\end{equation}
Since
\begin{equation}
\left[\frac{\partial  \Delta \vec A (\vec r,t)}{\partial t}\right]_{k} = (k+1)\Delta \vec A_{k+1}(\vec r)~,
\end{equation}
we can rewrite Eq.~(\ref{keyequation2}) in the following form:
\begin{eqnarray}\label{keyequation3}
n_0(\vec r)  (l+1) \Delta \vec A_{l+1}(\vec r)&=& -\sum_{k=0}^{l-1}n_{l-k}(\vec r)(k+1) \Delta \vec A_{k+1}(\vec r)  +  \sum_{k=0}^{l}\left \{\vec j_{l-k}(\vec r)\times \left[\vec \nabla \times \Delta \vec A_k(\vec r)\right] \right\}\nonumber \\ &+& [{\bf \nabla}\cdot \langle{\bf \hat \sigma}(\vec r,t)\rangle^\prime]_l - [{\bf \nabla} \cdot \langle{\bf \hat \sigma'}(\vec r,t)\rangle]_l~,\nonumber\\
\end{eqnarray}
where we have isolated on the left hand side the $k=l$ term of the sum which originally appeared on the left hand side of Eq.~(\ref{keyequation2}).  

We now show that Eq.~(\ref{keyequation3}) is effectively a recursion relation for the coefficients of the Taylor expansion of $\Delta \vec A(\vec r,t)$,  i.e., a relation that expresses $\Delta \vec A_{l+1}(\vec r)$ in terms of  $\Delta \vec A_{k}(\vec r)$, with $k \leq l$.   To this end, we must show that the right hand side of Eq.~(\ref{keyequation3}) depends only on coefficient $\Delta \vec A_{k}(\vec r)$, with $k \leq l$.  This is obviously true for the terms in which $\Delta \vec A_{k}$  appears explicitly.  There are also {\it implicit} $\Delta \vec A_{k}$s, which are hidden in the coefficients of the expansion of the density, the current density, and the expectation value of the stress tensor.   However, the structure of the time-dependent Schr\"odinger equation~(\ref{SE}), which is of first order in time,  guarantees the $l$-th coefficient in the Taylor expansion of the  quantum states $|\psi(t)\rangle$ and $|\psi'(t)\rangle$  is entirely determined by coefficients of order $k<l$ in the Taylor expansion of $\vec A$ and $\vec A'$:  hence all the quantities on the right hand side of Eq.~(\ref{keyequation3}) are completely determined by the coefficients $\Delta \vec A_k(\vec r)$, with $k\leq l$.  (In this argument $\vec A$ is considered a known quantity, and $\vec A' = \vec A +\Delta \vec A$ is the quantity we are trying to determine).

At this point, in order to make the recursion relation~(\ref{keyequation3}) work we only need to determine the  initial value of $\Delta \vec A$, namely $\Delta \vec A_0(\vec r) = \vec A'(\vec r,0)-\vec A(\vec r,0)$.  This is easily done, since from the equality of the densities and current densities of the primed and unprimed systems it follows that
\begin{equation}\label{initial}
n(\vec r,0)\Delta \vec A_0(\vec r)  = \langle \psi'(0)|\hat{\vec j}_p(\vec r)|\psi'(0) \rangle-  \langle \psi(0)| \hat{\vec j}_p(\vec r)|\psi(0) \rangle~,
\end{equation}
where $\hat{\vec j}_p (\vec r) \equiv \frac {1}{2m}\sum_i \left \{\hat{\vec p}_i, \delta (\vec r- \hat{\vec r}_i)   \right\}$  is the ``paramagnetic" current density operator, which has the same form in the primed and unprimed system.  Thus, the recursion relation~(\ref{keyequation3}), together with the initial condition~(\ref{initial}), completely determines the Taylor expansion of the potential $\vec A'(\vec r,t)$ that yields, in the primed system,  the same current density that $\vec A(\vec r,t)$ yields in the unprimed one.  According to our hypotheses,  a knowledge of the coefficients of the Taylor expansion of $\vec A'(\vec r,t)$ is equivalent to a knowledge of the function $\vec A'(\vec r,t)$ itself.  This completes the proof of our theorem.

\section{Discussion}
Two special cases of the theorem proved in the previous section deserve a special discussion.  
\begin{enumerate}
\item  The primed system coincides with the unprimed system, i.e. $U=U'$ and $|\psi(0)\rangle = |\psi'(0)\rangle$.  In this case Eq.~(\ref{initial}) above implies that $\Delta A_0(\vec r)=0$, and then it follows from Eq.~(\ref{keyequation3}) that $\Delta A_k(\vec r)=0$ for all $k$, i.e. $\vec A'(\vec r,t)=\vec A(\vec r,t)$ at all times.  This important result is just a statement of the Runge-Gross theorem for TDCDFT:  it asserts that two vector potentials that produce the same current density starting from the same initial state of a many-particle system must necessarily coincide, up to a gauge transformation: the map from vector potentials to current densities is invertible.  As noted in the introduction, this theorem was first proved by Ghosh and Dhara~\cite{Ghosh88} by a different method, similar in spirit to the original proof of the Runge-Gross theorem.  The present proof provides a simpler route to the same conclusion.

\item The primed system is non interacting, i.e., $U'=0$.  In this case our theorem provides a resolution of what could be called the non-interacting $\vec A$-representability problem.  In other words, the theorem shows that the current density produced by a vector potential $\vec A$ in an interacting many-particle system can also be obtained in a non-interacting system, under the action of a suitable vector potential $\vec A'$.  This is certainly possible if $|\psi'(0)\rangle=|\psi(0) \rangle$ (in which case we must have $\vec A'(\vec r,0)=\vec A(\vec r,0)$), but it may be more generally possible if one chooses for $|\psi'(0)\rangle$ a single Slater determinant (or even a noninteracting ground-state) that yields the correct initial density and current density.  
Thus, the theorem provides a solid basis for the use of the time-dependent Kohn-Sham equation, which indeed attempts to reproduce the correct current density in a system of noninteracting particles.   As pointed out in the introduction, this important result lays the foundation for the application of TDCDFT to molecular transport problems.  Notice, however, that the theorem does not say anything about the possibility of producing an {\it arbitrary} time-dependent current density by means of a suitable vector potential.
\end{enumerate}

\begin{acknowledgements}
The author acknowledges support from NSF Grant No.  DMR-0313681. Useful discussions with M. di Ventra, R. van Leeuwen, and Paul de Boeij are gratefully acknowledged. 
\end{acknowledgements}


\begin{thebibliography}{20}
\bibitem{footnote}{An additional hypothesis of the Runge-Gross theorem is that the quantity $n(\vec r,t) |\vec \nabla V^2(\vec r,t)|$ vanish at infinity  more rapidly than $\frac{1}{r^2}$ in three dimensions,  or satisfy periodic boundary conditions at infinity.}
\end{thebibliography}
\end{document}